\documentclass[lineno]{jfm}

\usepackage{graphicx}
\usepackage{gensymb}
\usepackage{subfigure}
\usepackage{float}
\usepackage{epstopdf}
\usepackage{epsfig}
\usepackage[figuresright]{rotating}
\usepackage{adjustbox}
\usepackage{makecell}
\usepackage{bm}
\usepackage{newtxtext}
\usepackage{newtxmath}
\usepackage{natbib}
\usepackage{hyperref}
\hypersetup{
	colorlinks = true,
	urlcolor   = red,
	citecolor  = blue,
}

\newcommand{\RomanNumeralCaps}[1]
\linenumbers

\title{NiST: a non-localized spatial-temporal constitutive relation in rarefied gas dynamics}

\author{Xiaoda Li, Bin Hu, \and Lei Wu
	\corresp{\email{wul@sustech.edu.cn}}
}

\affiliation{Department of Mechanics and Aerospace Engineering, Southern University of Science and Technology, Shenzhen 518055, China }
\begin{document}
	\maketitle
	
\begin{abstract}
Although the mesoscopic Boltzmann equation describes the rarefied gas dynamics, finding its solutions in complicated engineering problems is challenging. Therefore, over the past one and a half centuries, many partial differential equations based on a few  macroscopic variables are proposed. However, they not only have complicated forms, but also cannot make satisfactory prediction when the Knudsen number is large. Here, we propose a non-localized spatial-temporal (NiST) constitutive relation for rarefied gas dynamics, where the stress/heat flux at time $t$ and position $\bm x$ is determined by the velocity/temperature gradient in the nearby spatial-temporal coordinates, via convolution operators. By using the solutions of the Boltzmann equation for the Couette/Fourier flow and the spontaneous Rayleigh-Brillouin scattering, we extract the universal parameters of non-locality as functions of the spatial and temporal Knudsen numbers. Further tests in the sound propagation in rarefied gas show that the NiST constitutive relation can predict the rarefied gas flow over a wide range of Knudsen number.
\end{abstract}
		
\section{Introduction} \label{sec:introduction}
 
One of the fundamental tasks in the study of mechanics is to find the constitutive relations in the framework of continuum mechanics. For instances, in solid mechanics, one has to unveil the relation between the stress and strain, which is so far still the major challenge for complex materials~\citep{Eringen1976,cite1}. In the heat transfer in small-scale solid devices, the Fourier law predicts an infinite speed of heat conduction, which violates the relativity and goes against the experimental observations of heat waves; hence many hyperbolic equations are proposed to replace the parabolic heat conduction equation~\citep{cite4}.  In fluid mechanics, although it is believed that the Navier-Stokes-Fourier (NSF) equations can be used to describe turbulence, the huge computational cost at high Reynolds numbers calls for cheap and accurate turbulence models~\citep{Wilcox}, which encapsulate the constitutive relation between the Reynolds stress and velocity gradient. The same problem occurs in 
rarefied gas dynamics: while the mesoscopic Boltzmann equation is the fundamental equation for dilute gas flow from the continuum to free-molecular regimes, the high computational cost prohibits its wide range engineering applications; although many macroscopic equations derived from the Boltzmann equation by the Hilbert expansion~\citep{Hilbert1912,Sone2002Book}, the Chapman-Enskog expansion~\citep{Chapman1916,Enskog1917,CE}, and the Grad moment method~\citep{Grad1949,henning,TorrihonReview2016} are used describe the constitutive relations beyond the Newton law of viscosity and the Fourier law of thermal conductivity, their complexity and limited range of application~\citep{cite8} also call for new concise and universal constitutive relations.

The gas rarefaction effects occur when the Knudsen number is moderate and large.   In fact, there are two Knudsen numbers~\citep{Lei2022}. The first one is the spatial Knudsen number, which is defined as the ratio of the mean free path $\lambda$ of the gas molecules to the characteristic length $L$ of the flow:
\begin{equation}\label{Kn_original}
	\text{Kn}=\frac{\lambda}{L}, \quad\text{with}\quad
	\lambda=\frac{\mu}{p}\sqrt{\frac{\pi{}k_BT_0}{2m}},
\end{equation}
where $\mu$ is the shear viscosity, $p$ is the gas pressure, $k_B$ is the Boltzmann constant, $T_0$ is the gas temperature, and $m$ is the molecular mass. The second one is the temporal Knudsen number, which is defined as the ratio of the characteristic flow frequency $\omega$ to the mean collision frequency $1/\tau_c=p/\mu$ of gas molecules:
\begin{equation}\label{Kn_temporal}
	\text{Kn}_t=\omega\frac{\mu}{p}.
\end{equation}
When $Kn, Kn_t\lesssim0.1$, the gas flow is in the slip regime, where the NSF equations with the velocity-slip and temperature-jump boundary conditions may still be valid. When $0.1\lesssim Kn, Kn_t\lesssim10$ and $10\lesssim Kn, Kn_t$, the gas flows are in the transition and free-molecular regimes, respectively, where the traditional NSF equations cannot be applied. The primary reason is that, as the Knudsen number increases,  physical quantities at the point $\bm{x}$ and/or time $t$ will be influenced by those in the surrounding areas which is a few mean free path and/or mean free collision time away, according to the streaming and collision nature of the Boltzmann equation.

% because its theoretical basis has been destroyed

%Let's recall that, in solid mechanics, the hysteresis, relaxation, and creep are common phenomena in many materials. Maxwell, Kelvin, and Voigt combined spring and damper models to describe these viscoelastic behaviors. Consider a simple rod with an elongation of $u(t)$ under the action of force $F(t)$ in the one-dimensional case.  Its elongation is caused by the total history of loading from the start time to time $t$. If function $F(t)$ is continuous and differentiable, then the increment loaded within a small time interval $\tau$ is $(dF/d\tau)d\tau$. This increment continuously acts on the rod and contributes an element $du(t)$ to the elongation at time $t$, with its proportional creep function $c$ depending on the time interval $t-\tau$. So, as summarized in the Boltzmann superposition principle, the elongation of a simple rod under force is caused by the total history of loading from the beginning to the target time:
%\begin{equation}\label{nonlocal_stress_strain}
%	u(t) = \int_0^t c(t-\tau)\frac{dF(\tau)}{d\tau} d\tau.
%\end{equation}
%This is obviously a case of temporal nonlocality in the stress-strain relation, which is allowed in continuum mechanics, since, as defined by Fung~\citep{cite9}, the basic assumption in continuum mechanics is that ``the stress at a point is related to the strain and the rate of change of strain with respect to time at the same point'', i.e., the constitutive relation in continuum mechanics can be temporally nonlocal but spatially local.

Therefore, it is natural to assume that the stress and heat flux are related to the velocity and temperature gradients at neighbouring spatial-temporal coordinates. In fact, such assumptions are used in heat transfer and solid mechanics.
In the field of heat transfer, the Fourier law states that the heat flux $\bm q$ is proportional to the temperature gradient $-\nabla T$, with the proportionality constant being the thermal conductivity $\kappa$. This would lead to a parabolic equation for temperature evolution which has infinite propagation speed. \cite{cite4} added a time relaxation term to convert the original parabolic equation to a hyperbolic equation that has finite propagation speed, and eventually the heat flux is expressed in terms of the following convolution:
\begin{equation}
	\bm q(\bm x,t) = -\kappa \nabla \int_0^t G(|t-t'|) \frac{d T(\bm x,t')}{d t'} dt',
\end{equation}
where the kernel function is $G(t) = \tau_{lag}^{-1} \exp(-t/\tau_{lag})$. When the oscillation period of the thermal wave is much larger than the lagging time $\tau_{lag}$, e.g., in the limit of steady-state heat conduction, the kernel function degenerates to the delta function, so that the Fourier law is recovered. In other cases, it has a nonlocal memory effect. Similarly, in solid mechanics, Eringen believed that the stress $\sigma_{ij}(\boldsymbol{x})$ in solids depends on not only the strain $e_{ij}$ at the reference point $\boldsymbol{x}$, but also the strain at all points around that point~\citep{Eringen1976,cite1}:
\begin{equation}\label{eq5}
	\sigma_{ij}(\boldsymbol{x},t) = \int_V K(|\boldsymbol{x'}-\boldsymbol{x}|)C_{ijkl}e_{kl}(\boldsymbol{x'},t)d\boldsymbol{x'},
\end{equation}
where $C_{ijkl}$ is the elastic modulus tensor of classical isotropic elasticity. 
It has been found that such spatial nonlocalized constitutive relation can provide a powerful insight into the wave propagation under small-scale effects~\citep{cite2}, while in large solid the theory is reduced to the traditional linear elasticity because the kennel function $G$ degenerates to the delta function.
Such nonlocal linear elasticity has  been applied to explore the wave characteristics, dispersion relation, spiral dislocation, and edge dislocation in solids~\citep{cite1,cite11}.

In this paper, we shall combine the spatial and temporal non-localities together to construct a non-localized spatial-temporal (NiST) constitutive relation for rarefied gas flows, with very little modification to the traditional NSF equations. This combination will lead to the modification of transport coefficients (i.e., viscosity and thermal conductivity) so that it outperforms the traditional NSF equations, as well as the moment equations~\citep{cite8}; also, the resulting governing equation is linearly stable.

%The rest of the paper is organized as follows.  In Section~\ref{generalForm} we propose the general forms of the NiST for rarefied gas flows, while in Section~\ref{dataExtraction} we determine the model parameters  in the convolution kernels by comparing the NiST solutions for the Couette/Fourier flow and spontaneous Rayleigh-Brillouin scattering problem with those obtained from the linearized Boltzmann equation. In Section~\ref{univesal}, we determine the universal NiST models and test its accuracy in the Poiseuille flow and sound propagation in rarefied gas. Finally, conclusions are given in Section~\ref{conclusion}.

\section{{NiST} in rarefied gas flows}\label{generalForm}
	
To demonstrate the essential physics but make the formula simple, here we consider only the linearized flows of monatomic gas. The macroscopic equations from the conservation laws of mass, momentum and energy can be established as:
	\begin{equation}\label{eq7}
		\begin{aligned}[b]
			\frac{\partial \rho}{\partial t} + \frac{\partial u_j}{\partial x_j} = 0, \quad
			2\frac{\partial u_i}{\partial t} + \frac{\partial \rho}{\partial x_i} + \frac{\partial T}{\partial x_i} + \frac{\partial \sigma_{ij}}{\partial x_j}= 0, \quad
			\frac{3}{2}\frac{\partial T}{\partial t}+\frac{\partial u_j}{\partial x_j}
			+\frac{\partial q_j}{\partial x_j}=0,
		\end{aligned}
	\end{equation}
where $\rho$ is the perturbation mass density normalized by the average density $\rho_0$, $T$ is the perturbation temperature normalized by the average temperature $T_0$, $u$ is the flow velocity normalized by the most probable speed $v_m=\sqrt{2k_BT_0/m}$, $x$ and $t$ are the spatial and temporal variables normalized by the characteristic length $L_0$ and the characteristic time $L_0/v_m$, respectively; $\sigma_{ij}$ is the stress tensor normalized by the average pressure $p_0=\rho_0k_B T_0/m$, and $q_j$ is the heat flux normalized by $p_0v_m$. Note that the subscripts $i$ and $j$ represents the three orthogonal coordinates in the Cartesian coordinates, and the Einstein summation is used in the above equations.  % $a_i$ is the acceleration normalized by $v_m^2/L_0$;  
	
To close these equations, we need to know the constitutive relations, i.e., to express the stress and heat flux in terms of the flow density, velocity and temperature. In the continuum flow regime, through experiment, it is found that the stress tensor and heat flux can be expressed as (suppose the Prandtl number is $2/3$): 
\begin{equation}\label{eq11}
	\begin{aligned}[b]
		\sigma_{ij}^{NSF}  = -\text{Kn}'\left(\frac{\partial u_i}{\partial x_j}+\frac{\partial u_j}{\partial x_i}-\frac{2}{3}\frac{\partial u_k}{\partial x_k}\delta_{ij}\right),\quad
		q_i^{NSF} = -\frac{15}{8}\text{Kn}' \frac{\partial T}{\partial x_i}, 
	\end{aligned}
\end{equation}
where $\text{Kn}'=2\text{Kn}/{\sqrt{\pi}}$, and $\delta_{ij}$ is the Kronecker delta.

It should be noted that, these constitutive relations can also be derived from the Boltzmann equation via the Chapman-Enskog expansion~\citep{CE}, when the Knudsen number is small. 
When the Knudsen number is large, higher-order expressions for the stress and heat flux can also be derived, such as those in the  Burnett and super-Burnett equations, which are accurate to the second- and third-order of the Knudsen number, respectively. However, these higher-order constitutive relations are linearly unstable~\citep{Bobylev2006,GarciaColin2008} and hence they are rarely used nowadays.
	
In order to continue the use of macroscopic equations in rarefied flow, many modifications to the NSF equations are introduced.
On the one hand, the velocity-slip and temperature-jump boundary conditions are used, but they perform well only in the slip flow region~\citep{cite13}.  
On the other hand, the effective viscosity and thermal conductivity are introduced, based on the distance to the solid walls~\citep{zhang2005PRE,Dongari2012,Dongari2013PoF}. %Dongari2011,,Gijare2019, cite14,
However, in problems like the Rayleigh-Brillouin scattering where there are no solid wall involved, it is impossible to modify the transport coefficients to get better descriptions of gas rarefaction effects according to these theories~\citep{cite8}. Moreover, these methods did not take into account the effects of temporal non-locality.

Given that the ideas of nonlocality have been successfully applied to various fields such as the carbon nanotube and non-Fourier heat transfer in small-size solids~\citep{cite4,cite5,cite6,Eringen1976,cite1}, and the fact that the streaming operator in the Boltzmann equation is inherently nonlocal, we believe that the rarefaction effects can also be described using nonlocal constitutive relations. Therefore, we propose the following NiST constitutive relations:
\begin{eqnarray}
	\sigma_{ij}(x,t) = &2\int_{0}^t \int_{-\infty}^{\infty} K(|x'-x|,\ell_1) G(|t'-t|,\tau_{1})\sigma_{ij}^{NSF}(x',t') dx'dt', \label{eq13} \\
	q_{i}(x,t) = &2\int_{0}^t \int_{-\infty}^{\infty}  K(|x'-x|,\ell_2) G(|t'-t|,\tau_{2})q_{i}^{NSF}(x',t') dx'dt', \label{eq14}
\end{eqnarray}
where the parameters $\ell$ and $\tau$, as measures of the spatial and temporal nonlocality,  have been normalized by $L_0$ and $L_0/v_m$, respectively. As will be shown later, this NiST is able to capture the effective transport coefficients not only in the vicinity of solid walls, but also in the bulk flow. Also, the temporal nonlocality is formally taken into account, which is overlooked in previous works.
	
The spatial kernel function $K$ has the following properties~\citep{cite1}. First, 
it satisfies the normalization condition 
$\int K(|x'-x|,\ell) dx'= 1$.
Also, when $\ell \rightarrow 0$, $K$ must converge to the $\delta$ function, so that the nonlocal constitutive relations degenerate to the local one, i.e., the Newton law of viscosity. 
Second, the symmetry should be preserved with respect to $x$. Also, it acquires its maximum at $x'=x$ and has to decay to zero at large distances. Although other forms of $K$ might be used,
here we adopt the following bi-Helmholtz kernel~\citep{cite15}:
\begin{equation}\label{eq19}
	K(|x'-x|,\ell)= \frac{1}{2 \ell} \left(\frac{1}{\sqrt{2}}+\frac{|x'-x|}{\ell}\right)\exp\left(-\frac{\sqrt{2}|x'-x|}{\ell}\right).
\end{equation}
Special attention should be paid to wall-bounded problems, where the integration region in \eqref{eq13} and~\eqref{eq14} goes beyond the flow domain. To fix it some scholars introduced a local term $\delta(|x'-x|)$ to the original kernel function $K$, and obtained the following new kernel function~\citep{cite15}:
\begin{equation}\label{eq20}
	% (|x'-x|,\tau_1)
	K' = k_1 \delta(|x'-x|) + k_2 K(|x'-x|,\ell),
\end{equation}
where $k_1 = 1-\int_x K(|x'-x|,\ell)d{x}$ and $k_2 = 1-k_1$. %This form is also used in the NiST constitutive relations for wall-bounded flows in this paper.

We choose the temporal kernel function $G$ to be exactly the same as $K$, but with $x$ replaced by $t$, and $\ell$ replaced by $\tau$. Also note that $G$ works when $t'$ is in the interval of $(0,t)$ in \eqref{eq13} and \eqref{eq14}.

\section{Determination of NiST parameters}~\label{dataExtraction}

In this section, we shall use the solutions of the Couette/Fourier flow, as well as the solution of the spontaneous Rayleigh-Brillouin scattering, to determine the non-locality parameters $\ell_1, \ell_2, \tau_1, \tau_2$ in \eqref{eq13} and \eqref{eq14}. 

\subsection{Couette flow}

Consider the steady Couette flow between two plates located parallel to the  $x$ axis. The plate at $y=0$ is stationary, while the one at  $y=H$ is moving with a velocity $u_{wall}$ in the $x$ direction. $u_{wall}$ is small when compared to the most probable speed $v_m$ so that the problem can be linearized. The direction perpendicular to the flow is the $y$ direction. So we have $u_y = 0,~ \frac{\partial}{\partial t} = 0,~ q_i = 0$, and the density and temperature are constant. Choosing the characteristic length as $L_0=H$,
the constitutive relation~\eqref{eq13} is simplified to 
\begin{equation}\label{eq23}
	\sigma_{yx}(y) = -\text{Kn}' \int K'(| y - y'|,\ell_1) \frac{\partial u_x}{\partial y}\bigg|_{y'}  d y' \equiv \text{const}.
\end{equation}
In this case, the spatial nonlocality, due to the presence of solid wall, will generate an effective normalized shear viscosity ${\mu}_{e}$ which is a function of the wall distance:
\begin{equation}\label{eq24}
	{\mu}_{e}(y) =-\frac{{\sigma}_{yx}}{\text{Kn}'} \left(\frac{\partial {u}_x}{\partial {y}}\right)^{-1},
\end{equation}
and hence the velocity profile is not linear any more across the whole domain. 
The first-order slip boundary condition is
\begin{equation}\label{eq26}
	{u}-{u}_{\text{wall}} =-0.75\times \text{Kn}' \times {\mu}_{e}\big|_\text{wall} \times \frac{\partial {u}_x}{\partial {y}}\bigg|_\text{wall}.
\end{equation}

\begin{figure}
	\centering
	{\includegraphics[width=0.4\textwidth]{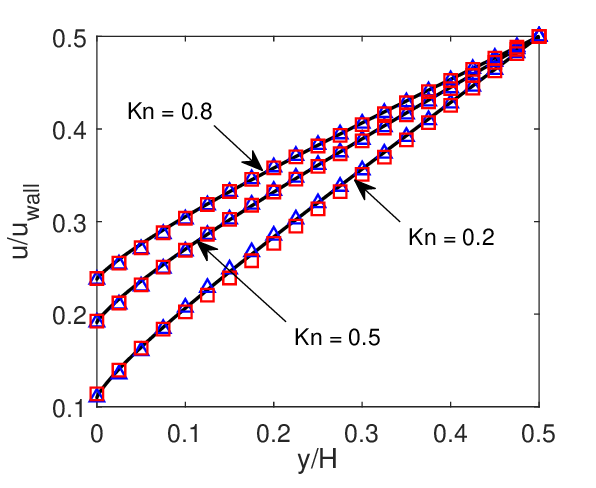} }
	{\includegraphics[width=0.4\textwidth]{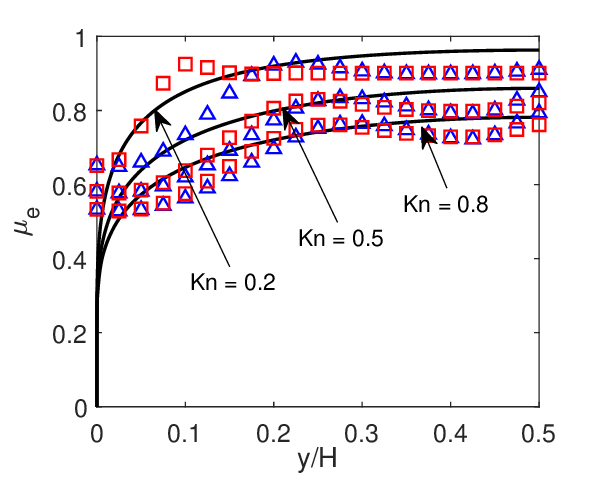} }
	\caption{Comparisons of the NiST results with that from the linearized Shakhov model in the Couette flow, in terms of the flow velocity and effective viscosity. Triangles and squares: results from the NiST model with $\ell_1$ given by \eqref{eq262} and \eqref{eq41}, respectively. Solid lines:  results from the linearized Shakhov model.
	}
	\label{resultc}
\end{figure}

\newpage 
The velocity profile obtained from the linearized Shakhov equation (a simplified Boltzmann equation) with the diffuse boundary condition is used as a norm to extract the parameter $\ell_1$~\citep{Lei2022}. That is, at each Knudsen number, $\ell_1$ is determined when the solution of \eqref{eq23} and \eqref{eq26} is closest to that of the Shakhov equation; figure~\ref{resultc} shows that the NiST model with proper value of $\ell_1$ yields a very good velocity profile when compared to that of the Shakhov equation. Eventually, the non-locality parameter $\ell_1$ can be fitted as the following function of the Knudsen number $\text{Kn}$:
\begin{equation}\label{eq262}
\ell_1 = \frac{0.3 \text{Kn} - 0.025 \text{Kn}^2}{0.12+\text{Kn}},
\end{equation}
which goes to zero as $\text{Kn}$ approaches zero. That is to say, when $\text{Kn}\rightarrow0$, $\ell_1\rightarrow0$, $K'(|y - y'|,\ell_1)\rightarrow\delta(y-y')$, and the constitutive relation in \eqref{eq23} approaches the Newton law of viscosity. 

We also compare the effective viscosity form the NiST model with that obtained from the Shakhov model in figure~\ref{resultc}. Due to the divergence of velocity gradient at the solid wall~\citep{takata2022JFM}, the effective viscosity from the Shakhov model is zero at the wall. This cannot be recovered from the NiST model. Nevertheless, away from the solid wall, the NiST model predicts a fairly good effective viscosity. By the way, if the Newton law of viscosity is used, the effective viscosity is always one, while that of the NiST model and Shakhov kinetic equation has reduced to about 0.7 when Kn=0.8.

\subsection{Fourier flow}

Consider the steady Fourier flow between two stationary parallel plates, the one at $y=H$ has a temperature $T_0+\Delta T/2$, while the one at $y=0$ has a temperature $T_0-\Delta T/2$. The temperature difference $\Delta T$ is small when compared to $T_0$, so that the problem can be linearized. Choosing the characteristic length as $L_0=H$, the constitutive relation~\eqref{eq14} is simplified to 
\begin{equation}\label{eq30}
	q_{y}(y) = - \frac{15}{8} \text{Kn}' \int K'(| y - y'|,\ell_2) \frac{\partial T}{\partial y}\bigg|_{y'}  d y' \equiv \text{const}.
\end{equation}
Similarly, due to the wall confinement, the effective normalized thermal conductivity $\kappa_e$ is a function of the wall distance:
\begin{equation}\label{eq243}
	{\kappa}_{e}(y) =- \frac{8}{15} \frac{{q}_y}{\text{Kn}'} \left(\frac{\partial {T}}{\partial {y}}\right)^{-1}.
\end{equation}
The first-order temperature-jump boundary condition is
\begin{equation}\label{eq30.5}
	T-T_{\text{wall}} = -1.33 \times \text{Kn}' \times \kappa_{e} \bigg|_\text{wall} \times \frac{\partial T}{\partial y}\bigg|_\text{wall}.
\end{equation}

Similar to the Couette flow, we can obtain the parameter $\ell_2$ by comparing the NiST solutions with those from the linearized Shakhov equation~\citep{Lei2022}:
\begin{equation}\label{eq263}
\ell_2 = \frac{0.3871 \text{Kn} - 0.1763 \text{Kn}^2+0.08 \text{Kn}^3 }{0.09548+\text{Kn}}.
\end{equation}
Clearly, when $\text{Kn}\rightarrow0$, $\ell_2\rightarrow0$, $K'(|y - y'|,\ell_2)\rightarrow\delta(y-y')$, and the constitutive relation in \eqref{eq30} approaches the Fourier law of thermal conductivity. 

\begin{figure}
	\centering
	{\includegraphics[width=0.4\textwidth]{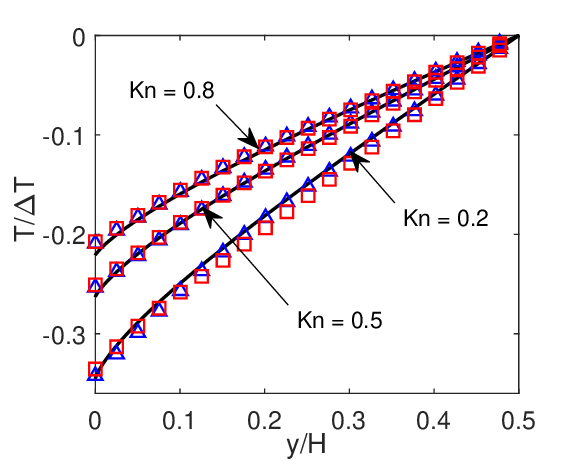}} 
	{\includegraphics[width=0.4\textwidth]{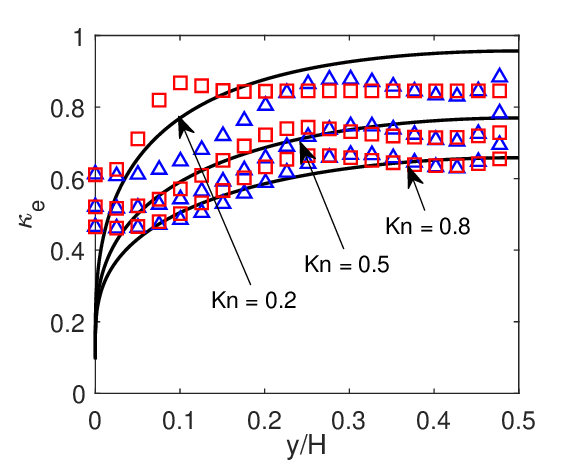}} 
	\caption{Comparisons of the NiST results with that from the linearized Shakhov kinetic model in the Fourier flow, in terms of the perturbed temperature and effective thermal conductivity. Triangles and squares: results from the NiST model with $\ell_2$ given by \eqref{eq263} and \eqref{eq41}, respectively. Solid lines: results from the linearized Shakhov model.
	}
	\label{resultf}
\end{figure}

It can be seen from figure~\ref{resultf} that, the NiST model predicts the temperature profile well. Also, although there are discrepancies in the effective thermal conductivity, the prediction accuracy is generally acceptable. By the way, if the Fourier law is used, the effective thermal conductivity is always one, while that of the NiST model and Shakhov equation has reduced to about 0.6 when Kn=0.8.

\subsection{Spontaneous Rayleigh-Brillouin scattering}

In order to make the NiST model applicable in unsteady flows, we use the Rayleigh-Brillouin scattering problem, where the laser light is scattered by the spontaneous fluctuation in gas density, to determine the temporal non-locality parameters.
In this case, both the spatial and temporal Knudsen numbers are important, and dozens of macroscopic equations fail to capture the spectrum up to $\text{Kn}=0.1$~\citep{cite8}. 

The characteristic length $L_0$ is chosen as the scattering wavelength, so that after the Laplace transform in $t$ and the Fourier transform in $x$, \eqref{eq7} is turned into the following form:
	\begin{equation}\label{eq37}
		\begin{bmatrix}
			-i\omega & 2 i \pi & 0 \\
			2 i \pi &  -2 i \omega + \frac{16}{3} \pi^2 \text{Kn} K^* (\ell_1)G^*(\tau_1) & 2 i \pi \\
			0 & 2 i \pi & -\frac{3}{2} i \omega + \frac{15}{2}  \pi^2 \text{Kn} K^*(\ell_2) G^*(\tau_2)
		\end{bmatrix}
		\begin{bmatrix}
			\hat{\rho}\\
			\hat{u}\\
			\hat{T}
		\end{bmatrix}
		=
		\begin{bmatrix}
			1\\
			0\\
			0
		\end{bmatrix},
	\end{equation}
where $i$ is the imaginary unit, $\omega$  is the  angular frequency normalized by $v_m/L_0$; $\hat{\rho}$, $\hat{u}$ and $\hat{T}$ are the spectra of the density, velocity, and temperature, respectively; and  
\begin{equation}\label{Fourier_G}
	\begin{aligned}[b]
		K^*(\ell) =\frac{4}{(2\pi\ell)^4+4(2\pi\ell)^2+4},\quad 
		G^*(\tau) &=\frac{4}{(\omega\tau )^4+4(\omega\tau)^2+4}.
	\end{aligned}
\end{equation}
It should be emphasized that both the spatial and temporal non-localities modify the effective viscosity and thermal conductivity from $\mu$ and $\kappa$ to $K^* (\ell_1)G^*(\tau_1)\mu$ and $K^*(\ell_2) G^*(\tau_2)\kappa$, respectively. This cannot be achieved when the effective transport coefficients are functions of the wall distance~\citep{zhang2005PRE,Dongari2012,Dongari2013PoF}.

The spectrum of the spontaneous Rayleigh-Brillouin scattering is given by the real part of $\hat{\rho}$, which can be calculated easily by linear algebra. And after optimization, the parameters $\ell_1, \ell_2, \tau_1, \tau_2$ are expressed in terms of the spatial and temporal Knudsen numbers as
 \begin{equation}\label{eq410}
 \begin{aligned}[b]
 		\ell=\ell_{1} &= \ell_{2} = 0.4262\text{Kn}, \\
 		\omega\tau=\omega\tau_{1} &= \omega\tau_{2} = 0.76\left(e^{0.76\text{Kn}}-e^{-10.76\text{Kn}}\right).
 %  \tau=\tau_{1} &= \tau_{2} = 0.76\left(e^{0.76\text{Kn}}-e^{-10.76\text{Kn}}\right)\frac{\text{Kn}}{\text{Kn}_t}.
 	\end{aligned}
 \end{equation}
Note that after normalization, the temporal Knudsen number is now expressed as 
$\text{Kn}_t=\omega \text{Kn}'$. 

\begin{figure}
	\centering
	\includegraphics[width=0.7\textwidth]{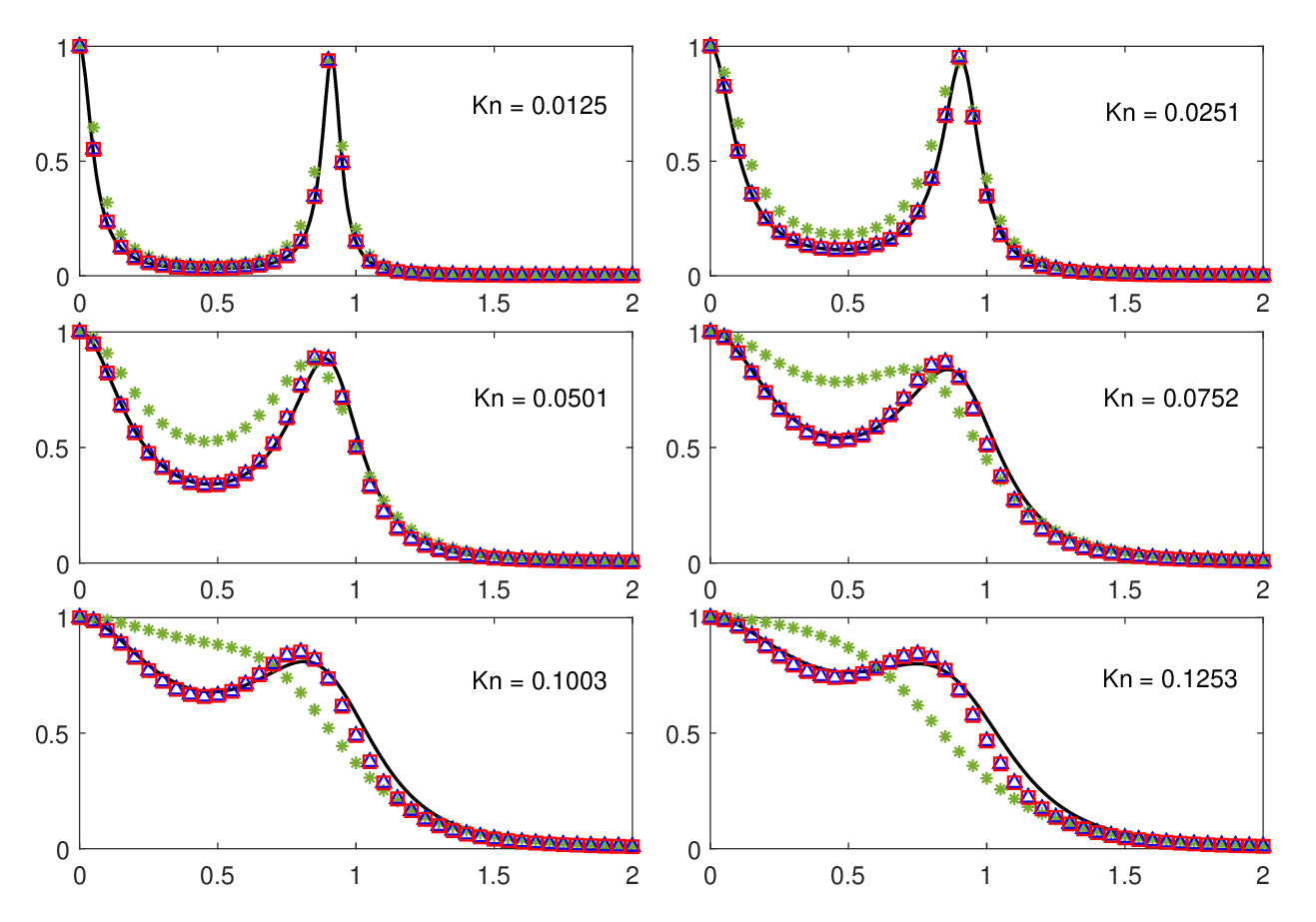}
	\caption{Spectra of the spontaneous Rayleigh-Brillouin scattering, as a function of the frequency $\omega/2\pi$. Stars and solid liens: results from the NSF equation and linearized Boltzmann equation, respectively. Triangles and squares: results from the NiST model with $\ell_2$ given by \eqref{eq410} and \eqref{eq41}, respectively. }
	\label{srbs}
\end{figure}

Figure~\ref{srbs} compares the results obtained from the NiST, NSF and linearized Boltzmann equation under different Knudsen number. While the NSF equations already have huge error when $\text{Kn}<0.05$, the NiST model is applicable  for the Knudsen number up to 0.125. Compared to the results in \cite{cite8}, the NiST model is not only better than the Burnett and super-Burnett equations which are respectively accurate to $O(\text{Kn}^2)$ and $O(\text{Kn}^3)$~\citep{GarciaColin2008}, but also better than the Grad 13, regularized 13 and 26 moment equations, which are accurate to $O(\text{Kn}^2)$, $O(\text{Kn}^3)$ and $O(\text{Kn}^5)$~\citep{henning,Gu2009JFM,TorrihonReview2016}, respectively.

\section{Universal parameters}\label{univesal}
 %This is somehow related to the physics-informed date-driven modeling \cite{ma2022dimensional}. 
 
In the previous section the parameters $\ell_1, \ell_2, 
\tau_1, \tau_2$ are optimized  case by case. Therefore, they might work on special flows. Here we attempt to obtain the universal expressions for these parameters, with a slight sacrifice in the prediction accuracy.

After comparison and optimization, the four parameters in the model can be expressed as
\begin{equation}\label{eq41}
	\begin{aligned}[b]
		\ell=\ell_{1} &= \ell_{2} = \min\left\{0.4262\text{Kn},0.1848\left(e^{0.2\text{Kn}}-e^{-10\text{Kn}}\right)\right\}, \\
  \omega\tau=\omega\tau_{1} &= \omega\tau_{2} = 0.76\left(e^{0.76\text{Kn}}-e^{-10.76\text{Kn}}\right).
		%\tau=\tau_{1} &= \tau_{2} = 0.76\left(e^{0.76\text{Kn}}-e^{-10.76\text{Kn}}\right)\frac{\text{Kn}}{\text{Kn}_t}.
	\end{aligned}
\end{equation}
We firstly apply the above parameters to the previous three problems. The results shown in figures  \ref{resultc}, \ref{resultf}, and \ref{srbs} as triangles confirm the universality and rationality of the universal parameters.
We have also checked that above NiST model can be applied to the Poiseuille flow; the results are not shown because the Couette flow and the Poiseuille flow are more or less the same.
%
%\subsection{Poiseuille flow}
%Poiseuille flow is a steady flow between two stationary parallel plates with the same temperature and is driven by pressure gradient $dp/dx$ (the direction of $p$ is along the $x$ axis which is parallel to these two plates). Combining the NiST model, Eq.~\eqref{eq7}, and the flow characteristics, we can obtain the governing equation of Poiseuille flow as Eq.~\eqref{eq50}
%\begin{equation}\label{eq50}
%\frac{\partial \sigma_{yx}}{\partial y} = \frac{dp}{dx}
%\end{equation}
%where, the $\sigma_{yx}$ is defined in Eq.~\eqref{eq23}. And the effective normalized shear viscosity in Poiseuille flow is same with that in Couette flow as Eq.~\eqref{eq24}. Finally, the corresponding boundary condition is
%\begin{equation}\label{eq500}
%	{u}-{u}_{\text{wall}} =-0.8\times \text{Kn}' \times {\mu}_{e}\big|_\text{wall} \times \frac{\partial {u}_x}       {\partial {y}}\bigg|_\text{wall}.
%\end{equation}
%

%\subsection{Sound propagation in rarefied gas}

% \begin{figure}
% 	\centering
% 	\includegraphics[width=0.85\textwidth]{resonator.png}
% 	\caption{ Schematic of the sound propagation in a resonator between transducer and receiver.}
% 	\label{resonator}
% \end{figure}

\begin{figure}
	\centering
 \includegraphics[width=0.5\textwidth]{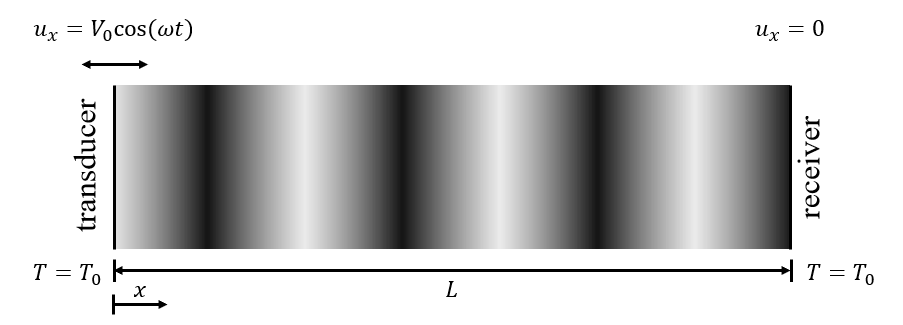}\\
	\includegraphics[width=0.85\textwidth]{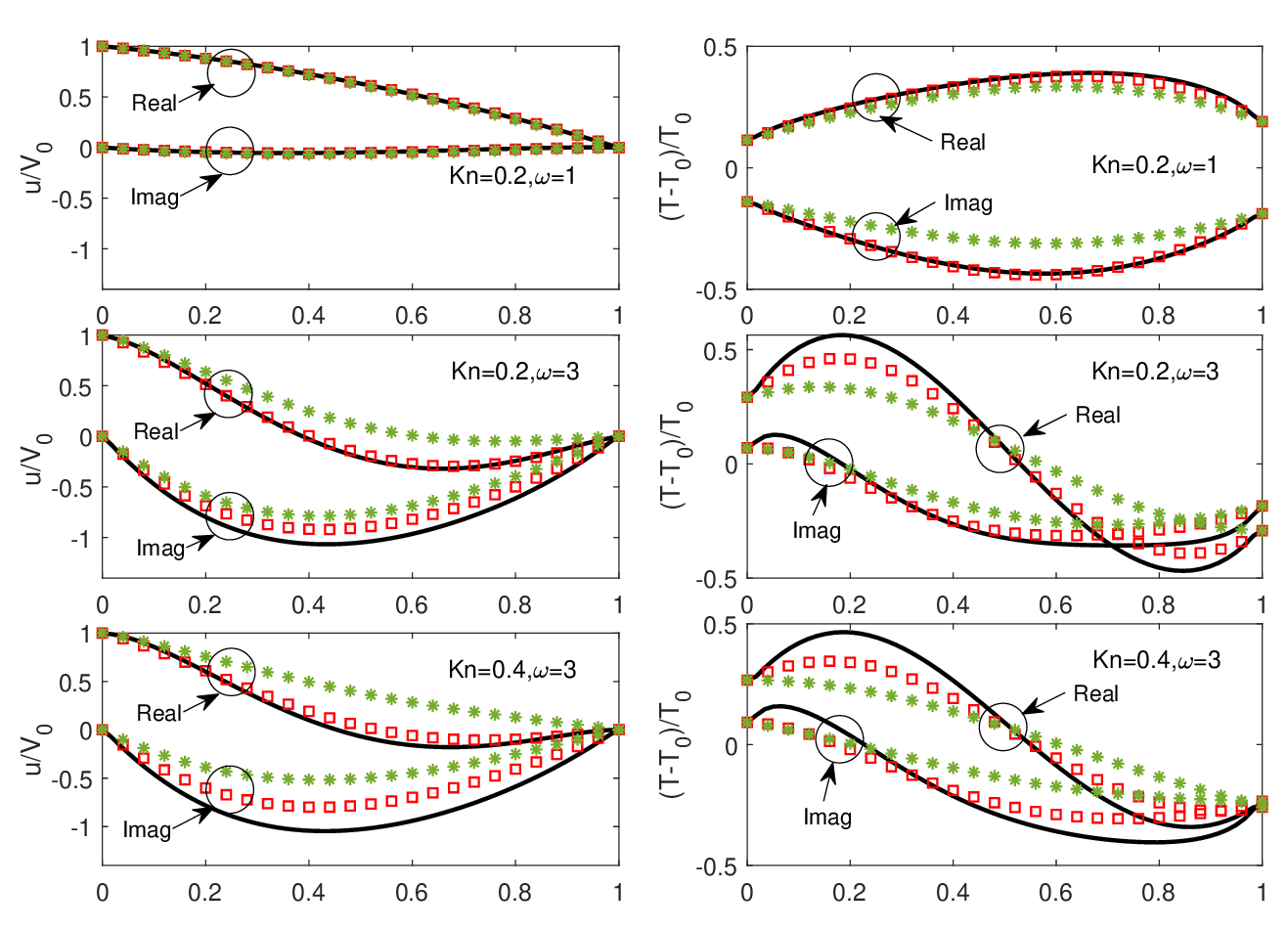}
	\caption{ The real and imaginary parts of the velocity and perturbed temperature profiles in the sound propagation problem, obtained from the NSF (star), NiST (square) and the Shakhov model (solid line). The horizontal coordinates are normalized distance.}
	\label{sw}
\end{figure}

We further assess the NiST model given by \eqref{eq13}, \eqref{eq14}, and \eqref{eq41} in the sound propagation in gas  confined between two parallel plates~\citep{cite18}. The left plate at $x=0$ is oscillating with fixed frequency which imposes the periodic velocity $u_x = V_0 \cos (\omega t)$, while the right plate at $x=L$ is stationary. And the wall temperature on both sides is equal to the average gas temperature $T_0$. Here, we first obtain the reference velocity and temperature profiles using the linearized Shakhov model. Utilizing the precise boundary condition provided by the Shakhov model, we then apply the NiST model and \eqref{eq7} to obtain the velocity of temperature profiles. To be specific, on top of the normalization described below \eqref{eq7}, all macroscopic quantities are further normalized by $V_0/v_m$, which is a small number so that the flow can be linearized. When the ``periodic steady-state'' is reached, all variables oscillate with the same frequency $\omega$. Therefore, we can apply the Fourier transform to~\eqref{eq7} to remove the time-dependence:
\begin{equation}\label{eq51}
	\begin{aligned}[b]
		i \omega \hat{\rho} =\frac{\partial \hat{u}}{\partial x}, \quad
		2 i \omega \hat{u}= \frac{\partial \hat{\rho}}{\partial x} + \frac{\partial \hat{T}}{\partial x} + G^\ast(\tau)\frac{\partial \hat{\sigma}_e}{\partial x}, \quad
		\frac{3}{2}i \omega \hat{T} = \frac{\partial \hat{u}}{\partial x} +G^\ast(\tau) \frac{\partial \hat{q}_e}{\partial x},
	\end{aligned}
\end{equation}
where $G^{\ast}(\tau)$ is given in \eqref{Fourier_G}, and
\begin{equation}
	\begin{aligned}[b]
		\hat{\sigma}_e =   2\int_{-\infty}^{\infty} K(|x'-x|,\ell) \sigma^{NSF}(x') dx',\quad
		\hat{q}_e =  2\int_{-\infty}^{\infty} K(|x'-x|,\ell) q^{NSF}(x') dx'.  
	\end{aligned}
\end{equation}

Comparisons in the velocity and temperature profiles, as obtained from the NSF equation, NiST model and the Shakhov model, are shown in figure~\ref{sw}. Note that here the macroscopic quantities are complex numbers, which contain the oscillation amplitude and the phase lag. When $\text{Kn}=0.2$ and $\omega=1$, the NiST model predicts the velocity and temperature very well, while the NSF equations only predict the velocity profile. Keeping the spatial Knudsen number fixed but increasing the angular frequency (equivalent to the increase of temporal Knudsen number), we see that the prediction power of the NSF equations and NiST model both decrease, but the NiST model decreases in a slower pace: the prediction error of the NiST model is relatively acceptable until $\text{Kn} = 0.3$ and $\omega = 3$, while the NSF equations only work up to $\text{Kn} = 0.1$ and $\omega = 1$. 
When $\text{Kn} = 0.4$ and $\omega = 3$, we see that the NiST model has some errors in the velocity and temperature profiles, although in the steady Fourier flow the NiST model can be applied as least up to $\text{Kn} = 1$. This is because in sound waves both the spatial and temporal Knudsen numbers are large, so that the applicability range will be reduced when only the spatial or temporal Knudsen number is large.

\section{Conclusions}\label{conclusion}
In summary, we have proposed a non-localized spatial-temporal constitutive relation to describe the rarefied gas dynamics, where the stress (heat flux) at the position $\bm{x}$ and time $t$ is related to velocity (temperature) gradient in the surrounding area which is a few mean free path away and the time history which is a few mean free collision time away. Therefore, the stress and heat flux are expressed in terms of the spatial-temporal convolutions, and the non-locality parameters in the convolution kernel have been extracted, independently, from the solutions of the Boltzmann model equation for the Couette flow, Fourier flow, and the Rayleigh-Brillouin scattering problem. Then, the universal non-locality parameters are proposed for general rarefied gas flows, which work well in the Couette/Fourier flow, spontaneous Rayleigh-Brillouin scattering, and the sound propagation in rarefied gas. Especially, the NiST model can predict the light scattering spectrum up to $\text{Kn}\approx 0.125$, which is better than the other dozens macroscopic equations (includes the regularized 26-moment equations which is derived from the Boltzmann equation and accurate to the fifth order of Kn).

Since the spatial-temporal convolutions modify only the effective viscosity and thermal conductivity, it can be shown that the proposed NiST model, together with the conservation equations for mass, momentum and energy, is linearly stable, and hence can be used in engineering applications to get quick solutions up to the transition flow regime. Finally, it should be noted that the present work focuses only on linear flows, as a proof of the NiST concept. In the future, we shall extend the NiST model to nonlinear rarefied gas flows. 

%in which the effective viscosity and thermal conductivity show similar distributions~\citep{Ou2020PoF,Myong1999PoF,Myong2001JCP}. So, it prompts us to think the possibility of the application of this idea in the nonlinear flows and this will be our future work.

\section{Acknowledgements}\label{sec:Acknowledgements}
	
This work is supported by the National Natural Science Foundation of China (12172162). 
	
\section{Declaration of interest}\label{sec:intrest}
	
The authors report no conflict of interest.

\bibliographystyle{jfm}
\bibliography{ref}
	
\end{document}